# Image potential of single-wall carbon nanotubes in the field emission condition

**Weiliang Wang, Jie Peng, Guihua Chen, Shaozhi Deng, Ningsheng Xu, Zhibing Li***

State Key Laboratory of Optoelectronic Materials and Technologies

School of Physics and Engineering, Sun Yat-Sen University, Guangzhou 510275, People's Republic of China

**Abstract:** *We calculated the image potentials of single-walled carbon nanotubes of various structures in the condition of field emission with a quantum chemistry method. The image potential of the single-walled carbon nanotubes can be well fitted with the image potential of an ideal metal sphere in a size comparable to an atom. The image potential is not sensitive to both the applied field and the structure of the tube. When the image potential is included, the emission current would be one order larger.*

In the field emission (FE) of conventional plane emitters, the image potential reduces the height of the potential barrier, and also can introduce image state at the surface, by which it enhances the field emission significantly. [1] The image potential of metal plane has been calculated with modified jellium model.[2] Their quantum mechanical results indicate that the classical model has over estimated the image potential. Edgcombe *et al.* have investigated the image potential of sharp emitters with the ideal metal model. [3] The image potential of nano-emitters, however, has not been fully clear. [4] For an emitter of nano scale, in principle, the occupation probability of the orbitals depends on both the testing electron and the applied field, thereby the image potential also. When the testing electron is close to the atoms of the nano-emitter, the correlation between the testing electron and the electrons would play role and that should be treated with the quantum mechanical many body theory. Since the jellium model can not be applied to the nano-emitters in which the effect of finite atom number is important, we appeal to more sophisticated methods. The DFT method [5] can provide detail electron properties of small systems of a few hundred atoms. In order to predict the charge distribution in a nano-emitter, for instance made of a single-walled carbon nanotube (SWCNT) that consists of hundred thousands of atoms, a multi-scaled method has been proposed.[6,7] With this method, the apex-vacuum electrostatic potential of SWCNTs of micrometer long has been calculated.

The image potential of SWCNTs of 1. μm long in the condition of field emission should be studied in the present paper. Our method has two steps. Firstly, we simulate the SWCNT without the testing charge by the multi-scale method [6,7] to obtain the charge distribution along the tube. Then, the tip subsystem that has the extra number of electrons found in the previous step is separated from the SWCNT, with the artificial open end saturated by hydrogen atoms, that is simulated by the DFT method [5] in the electric field of superposition external applied field, the field of the excessive charges in the rest of the tube, and the electric field of the testing charge. Each tip subsystem in our simulation consists about 100 atoms. This method has assumed that the testing charge would modify the electron distribution in the tip but would not change the total number of electrons in the tip. We investigated the open SWCNTs with chirality indexes (7, 0), (4, 4), (5, 5), and capped SWCNT with index (5, 5), with a testing point charge [8] put near the apex of the CNT as shown in Fig. 1 (plotted by MacMolPlt [9]). Then the electric field at the position of the testing point charge caused by the induced charges is calculated. With the electric fields for various testing charge positions along the straight line from the apex atom to anode in hand, one can obtain the potential barrier through a numerical integration of the electric fields along the line.

Fig. 2 shows the excessive charges at the apex of the (7, 0) SWCNT layer by layer. The layer labeled by zero is the top layer of the SWCNT. The charges are associated to atoms by the Mulliken method. [10] In Fig.2, the charges of the seven atoms on the same layer of the (7, 0) SWCNT are spread out in the interval of two nearby layer labels and are in fact belong to the layer with the smaller label. Fig.2 (a) and (b) are corresponding to applied fields 0 V/μm and 11 V/μm, respectively. To show the sensitivity of induced charges to the positions of the testing charge, two positions of the testing charge, i.e., z = 0.4 nm and z = 10. nm are presented in Fig.2. In the case of larger distance (z=10. nm), the effect of testing charge is in fact ignorable. In the case of z=0.4 nm, one can see clearly that the testing charge breaks the axial symmetry. It implies that the occupation probability of electronic orbitals is changed.

Fig. 3 is the corresponding apex-vacuum potential barrier with (solid) and without (dotted) image potential. The difference between the solid lines and their corresponding dotted lines is caused by the image charge, and is referred to as the image potential (Fig. 4). It is helpful to compare the image potentials of the SWCNTs with those of the ideal metal spheres and the ideal metal plane. The image potential of the ideal metal sphere of radii r is known as

$$W(a) = -\frac{e}{4\pi\varepsilon_0} \frac{r}{2(z^2 - r^2)} \tag{1}$$

When $r \to \infty$, one gets the image potential of the ideal metal plane

$$W(z) = -\frac{e}{4\pi\varepsilon_0} \frac{1}{4a} \tag{2}$$

where a = z – r is the distance between the testing point charge and the surface of the metal plane.

The image potential of the ideal metal sphere of r = 0.04 nm and 0.08 nm (close to the radii of an atom) and the image potential of the ideal metal plane are also shown in Fig. 4 for reference. Fig.4 shows that the image potentials of SWCNTs can be well approximated by the image potentials of the ideal metal spheres of those sizes, that incidentally agrees with the CNT model that the carbon atoms are represented by the metal spheres of r=0.07nm[11]. However, we found that the image potential is not only contributed by the atom that is nearest to the testing charge, but by other atoms also. The image potential of an isolated single carbon atom is represented by squares in Fig. 4, it is significantly smaller than that of the SWCNTs. Therefore, the image potential is not given only by the polarization of the individual atom but also related to the charge transfer between atoms. But the classical metal sphere model can not provide the correct apex potential barrier since it can not give the field penetration and charge distribution in the CNT. It should be stressed that the field penetration in the apex region has important contribution to the energy potential barrier for the emission, although incidentally one would fit the experimental observations only with the external applied field and the classical image potential [12]. That is because that the ignorance of the field penetration would be counteracted by an incorrect field enhancement factor that is usually much bigger than the actual one. Quantum simulations have shown that the field penetration can reduce the effective work function (potential barrier height) significantly [7].

The tunneling probability (and thus the emission current) from the potential barrier is calculated with the JWKB approximation. [13] Table I shows emission currents of the (7, 0), (4, 4), (5, 5) open and (5, 5) capped SWCNTs, with and without the image potential, under various applied fields.

To summary, we have calculated the image potentials of (7, 0), (4, 4), (5, 5) open SWCNTs and (5, 5) capped SWCNTs of 1. μm long in the condition of field emission with the quantum chemistry method. Our results show that the image potential is not sensitive to the applied fields, the emitting points, and the structures of SWCNTs. An unexpected result is that the open SWCNTs and the capped SWCNT have similar image potentials. The image potential can be well approximated by the image potential of the ideal metal sphere of an atom size. This observation provides a convenient way to include the image potential in the energy potential barrier that governs the emission probability. When the image potential is taken into account, the emission current could be one order larger than that the image potential is excluded.

ACKNOWLEDGMENTS

The project is supported by the National Natural Science Foundation of China (Grant Nos. 10674182, 90103028, and 90306016) and National Basic Research Program of China (973

program 2007CB935500).

*Table Ⅰ Emission current of (7, 0), (4, 4), (5, 5) open and (5, 5) capped SWCNTs with and without image potential under various applied fields.*

| SWCNT | (7, 0) | | (4, 4) | | (5, 5) open | | (5, 5) capped |
|---|---|---|---|---|---|---|---|
| Applied field (V/μm) | 11 | 15 | 7 | 13 | 7 | 11 | 12 |
| Emission current (μA) without image potential | $7.8 \times 10^{-4}$ | $1.4 \times 10^{-1}$ | $6.7 \times 10^{-18}$ | $8.1 \times 10^{-2}$ | $2.3 \times 10^{-18}$ | $1.4 \times 10^{-3}$ | $1.9 \times 10^{-5}$ |
| Emission current (μA) with image potential | $6.1 \times 10^{-3}$ | $6.1 \times 10^{-1}$ | $4.9 \times 10^{-17}$ | $5.2 \times 10^{-1}$ | $2.4 \times 10^{-17}$ | $1.2 \times 10^{-2}$ | $1.9 \times 10^{-4}$ |

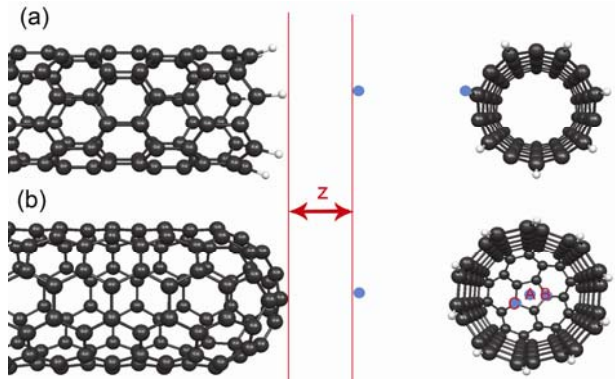

Fig. 1 (a) Schematic illustration of a testing point charge near the apex of the open SWCNT. (b) Same as (a), but for capped SWCNT, three possible emission sites are indicated. The black circles are carbon atoms and the white ones are hydrogen atoms, z is defined as the distance between the SWCNT and the testing point charge. (Left hand side: side view; right hand side: top view.)

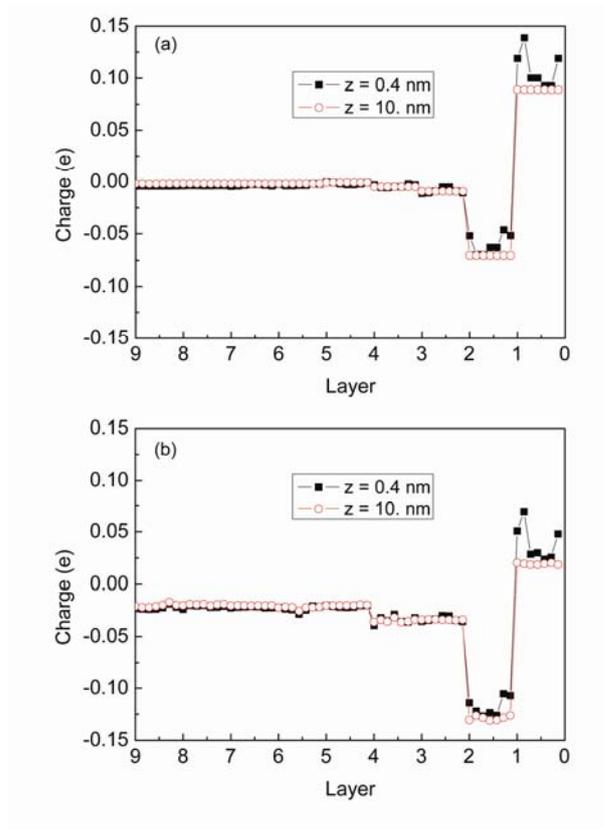

Fig. 2 Mulliken charge distribution at the apex of the (7, 0) SWCNT for $z = 0.4$ nm and $z = 10.$ nm under the applied field of (a) 0 V/μm and (b) 11 V/μm. The X coordinate is the number of atom layer. There are 7 atoms in each layer for (7, 0) SWCNT, the charges of atoms of the same layer are presented in the interval of two layer labels, in equal X spacing.

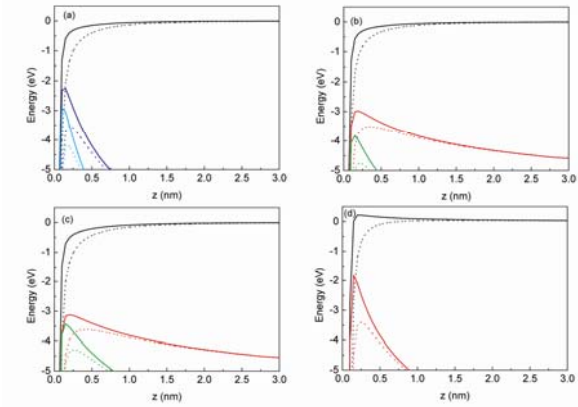

*Fig. 3 (a) Potential barriers of (7, 0) SWCNT with (solid) and without (dotted) image potential, from up to down corresponds to applied field of 0 V/μm, 11 V/μm, 15 V/μm. (b) Same as (a), but for (4, 4) SWCNT, from up to down corresponds to applied field of 0 V/μm, 7 V/μm, 13 V/μm. (c) Same as (a), but for (5, 5) SWCNT, from up to down corresponds to applied field of 0 V/μm, 7 V/μm, 11 V/μm. (d) Same as (a), but for capped (5, 5) SWCNT, from up to down corresponds to applied field of 0 V/μm, 12 V/μm, only potential barriers from emission site (C) are shown, that of site (A) and (B) are almost the same.*

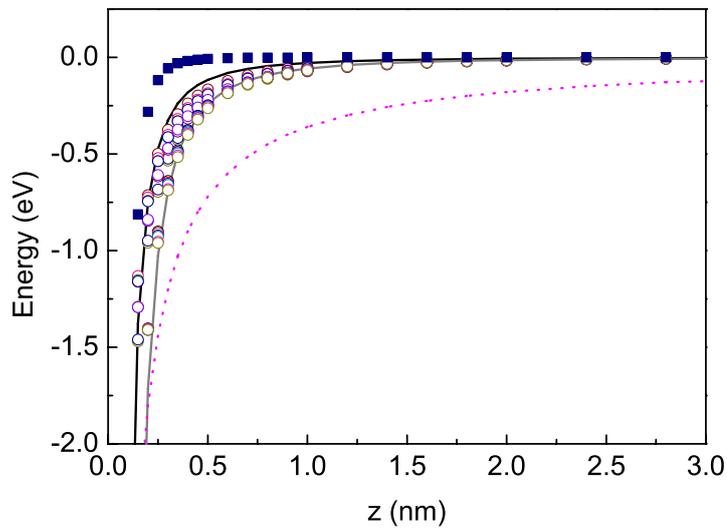

*Fig. 4 Image potentials of SWCNTs under various applied fields (open circyles), and image potential of ideal metal spheres of r = 0.04 nm (black line) and r = 0.08 nm (gray line) and of the plane (dotted line).*